\PassOptionsToPackage{unicode}{hyperref}
\PassOptionsToPackage{hyphens}{url}
\PassOptionsToPackage{dvipsnames,svgnames,x11names}{xcolor}
\documentclass[
  12pt]{article}

\usepackage{amsmath,amssymb}
\usepackage{iftex}
\ifPDFTeX
  \usepackage[T1]{fontenc}
  \usepackage[utf8]{inputenc}
  \usepackage{textcomp} 
\else 
  \usepackage{unicode-math}
  \defaultfontfeatures{Scale=MatchLowercase}
  \defaultfontfeatures[\rmfamily]{Ligatures=TeX,Scale=1}
\fi
\usepackage{lmodern}
\ifPDFTeX\else  
\fi
\IfFileExists{upquote.sty}{\usepackage{upquote}}{}
\IfFileExists{microtype.sty}{
  \usepackage[]{microtype}
  \UseMicrotypeSet[protrusion]{basicmath} 
}{}
\makeatletter
\@ifundefined{KOMAClassName}{
  \IfFileExists{parskip.sty}{%
    \usepackage{parskip}
  }{
    \setlength{\parindent}{0pt}
    \setlength{\parskip}{6pt plus 2pt minus 1pt}}
}{
  \KOMAoptions{parskip=half}}
\makeatother
\usepackage{xcolor}
\makeatletter
\ifx\paragraph\undefined\else
  \let\oldparagraph\paragraph
  \renewcommand{\paragraph}{
    \@ifstar
      \xxxParagraphStar
      \xxxParagraphNoStar
  }
  \newcommand{\xxxParagraphStar}[1]{\oldparagraph*{#1}\mbox{}}
  \newcommand{\xxxParagraphNoStar}[1]{\oldparagraph{#1}\mbox{}}
\fi
\ifx\subparagraph\undefined\else
  \let\oldsubparagraph\subparagraph
  \renewcommand{\subparagraph}{
    \@ifstar
      \xxxSubParagraphStar
      \xxxSubParagraphNoStar
  }
  \newcommand{\xxxSubParagraphStar}[1]{\oldsubparagraph*{#1}\mbox{}}
  \newcommand{\xxxSubParagraphNoStar}[1]{\oldsubparagraph{#1}\mbox{}}
\fi
\makeatother

\usepackage{longtable,booktabs,array}
\usepackage{calc} 
\usepackage{etoolbox}
\makeatletter
\patchcmd\longtable{\par}{\if@noskipsec\mbox{}\fi\par}{}{}
\makeatother
\IfFileExists{footnotehyper.sty}{\usepackage{footnotehyper}}{\usepackage{footnote}}
\makesavenoteenv{longtable}
\usepackage{graphicx}

\usepackage{tikz}
\usepackage{multicol} 
\usetikzlibrary{fit, arrows.meta, positioning}

\usepackage{amsthm}
\newtheorem{theorem}{Theorem}[section]
\newtheorem{proposition}[theorem]{Proposition}

\usepackage{algorithm} 
\usepackage{algpseudocode}

\usepackage{bm}

\usepackage{subfigure}

\makeatletter
\def\maxwidth{\ifdim\Gin@nat@width>\linewidth\linewidth\else\Gin@nat@width\fi}
\def\maxheight{\ifdim\Gin@nat@height>\textheight\textheight\else\Gin@nat@height\fi}
\makeatother
\setkeys{Gin}{width=\maxwidth,height=\maxheight,keepaspectratio}
\makeatletter
\def\fps@figure{htbp}
\makeatother

\makeatletter
\@ifpackageloaded{caption}{}{\usepackage{caption}}
\AtBeginDocument{%
\ifdefined\contentsname
  \renewcommand*\contentsname{Table of contents}
\else
  \newcommand\contentsname{Table of contents}
\fi
\ifdefined\listfigurename
  \renewcommand*\listfigurename{List of Figures}
\else
  \newcommand\listfigurename{List of Figures}
\fi
\ifdefined\listtablename
  \renewcommand*\listtablename{List of Tables}
\else
  \newcommand\listtablename{List of Tables}
\fi
\ifdefined\figurename
  \renewcommand*\figurename{Figure}
\else
  \newcommand\figurename{Figure}
\fi
\ifdefined\tablename
  \renewcommand*\tablename{Table}
\else
  \newcommand\tablename{Table}
\fi
}
\@ifpackageloaded{float}{}{\usepackage{float}}
\floatstyle{ruled}
\@ifundefined{c@chapter}{\newfloat{codelisting}{h}{lop}}{\newfloat{codelisting}{h}{lop}[chapter]}
\floatname{codelisting}{Listing}

\makeatother
\makeatletter
\@ifpackageloaded{caption}{}{\usepackage{caption}}
\@ifpackageloaded{subcaption}{}{\usepackage{subcaption}}
\makeatother

\ifLuaTeX
  \usepackage{selnolig}  
\fi
\usepackage[]{natbib}
\usepackage{bookmark}

\IfFileExists{xurl.sty}{\usepackage{xurl}}{} 
\hypersetup{
  pdftitle={Title},
  pdfauthor={Author 1; Author 2},
  pdfkeywords={3 to 6 keywords, that do not appear in the title},
  colorlinks=true,
  linkcolor={blue},
  filecolor={Maroon},
  citecolor={Blue},
  urlcolor={Blue},
  pdfcreator={LaTeX via pandoc}}

\newcommand{\anon}{0}

\begin{document}

\def\spacingset#1{\renewcommand{\baselinestretch}%
{#1}\small\normalsize} \spacingset{1}


\if0\anon
{
  \title{\bf Learning Heterogeneous Ordinal Graphical Models via Bayesian Nonparametric Clustering}
   \author{
    Wang Wen\textsuperscript{1}, 
    Ziqi Chen\textsuperscript{1}, 
    and Guanyu Hu\textsuperscript{2}
  }
  \date{
    \textsuperscript{1}KLATASDS-MOE, School of Statistics, East China Normal University, 
    Shanghai, 200062, China\\
    \textsuperscript{2}Department of Statistics \& Probability, Michigan State University, East Lansing, MI, 48824, USA\\   
  }

  \maketitle
} \fi

\if1\anon
{
  \bigskip
  \bigskip
  \bigskip
  \begin{center}
    {\LARGE\bf Learning Heterogeneous Ordinal Graphical Models via Bayesian Nonparametric Clustering}
\end{center}
  \medskip
} \fi

\bigskip
\begin{abstract}
Graphical models are powerful tools for capturing conditional dependence structures in complex systems but remain underexplored in analyzing ordinal data, especially in sports analytics. Ordinal variables, such as team rankings, player performance ratings, and survey responses, are pervasive in sports data but present unique challenges, particularly when accounting for heterogeneous subgroups, such as teams with varying styles or players with distinct roles. Existing methods, including probit graphical models, struggle with modeling heterogeneity and selecting the number of subgroups effectively. We propose a novel nonparametric Bayesian framework using the Mixture of Finite Mixtures (MFM) approach to address these challenges. Our method allows for flexible subgroup discovery and models each subgroup with a probit graphical model, simultaneously estimating the number of clusters and their configurations. We develop an efficient Gibbs sampling algorithm for inference, enabling robust estimation of cluster-specific structures and parameters. This framework is particularly suited to sports analytics, uncovering latent patterns in  player performance metrics. Our work bridges critical gaps in modeling ordinal data and provides a foundation for advanced decision-making in sports performance and strategy.

\end{abstract}

\noindent%
\vfill

\newpage
\spacingset{1.8} 

\section{Introduction}\label{sec-intro}
Graphical models provide a powerful probabilistic framework for exploring the conditional dependence structure of complex systems using graph theory \citep{lauritzen1996graphical}. These models have garnered significant attention in fields such as psychology \citep{fried2015loss,isvoranu2017network}, biology \citep{friedman2004inferring}, and medicine \citep{allen2012log}, due to their ability to uncover variable interactions. Among these, undirected graphical models are particularly prominent for modeling multivariate random variables. These models use undirected graphs where nodes represent variables, and edges denote conditional dependence, with the absence of an edge indicating conditional independence. Common examples include Gaussian graphical models for continuous data \citep{speed1986gaussian,rue2005gaussian}, Ising models for nominal data \citep{ising1925beitrag}, and hybrid models combining these types \citep{lauritzen1989graphical,yang2014mixed}.

Despite recent advancements, a significant gap remains in effectively modeling ordinal data, which is prevalent across many real-world applications. For example, ordinal scales are widely employed in medical research to categorize disease severity or progression stages, and in survey instruments to capture attitudes on a spectrum from ``strongly disagree'' to ``strongly agree''. These variables possess a natural ordering but lack precise numerical meaning, making traditional continuous models ill-suited for their analysis. This widespread occurrence underscores the need for graphical models specifically designed to handle ordinal variables, accounting for their distinctive characteristics and dependencies. Broadening the scope of graphical models in this direction not only enhances their practical utility but also facilitates richer and more accurate analyses of ordered data across diverse domains.

However, most existing ordinal graphical models assume a homogeneous population, which is often unrealistic in practice. Real-world datasets frequently consist of heterogeneous subgroups with distinct structures. For example, in psychopathological research, different sub-populations may exhibit varying symptom networks. Ignoring such heterogeneity can lead to misrepresentations of the underlying population structures \citep{brusco2019ising}. Addressing this limitation by incorporating methods to capture heterogeneous structures in ordinal data is critical for improving the accuracy and relevance of graphical models across diverse applications.
\subsection{Related Work and Challenges}
Graphical models for ordinal data have received considerably less attention than their continuous and nominal counterparts. A foundational approach in this domain is the multivariate probit model, which models the joint distribution of ordinal variables by introducing latent multivariate Gaussian variables. Building on the success of multivariate probit analysis \citep{albert1993bayesian, chib1998analysis}, \citet{guo2015graphical} proposed probit graphical models to capture dependence structures among ordinal variables. However, estimating sparse graphical structures within this framework remains challenging due to the intractability introduced by latent variables in the log-likelihood function. To overcome these difficulties, several alternative methods have been developed: \citet{guo2015graphical} proposed an approximate EM-like algorithm, \citet{suggala2017ordinal} introduced a two-stage procedure based on bivariate marginal log-likelihoods, and \citet{feng2019high} developed a rank-based ensemble estimation approach. More recently, \citet{castelletti2024learning} designed a Bayesian semi-parametric method for structure learning in directed networks that is also applicable to ordinal data. 

Graphical models have also shown promise in sports data analysis, particularly when dealing with ordinal outcomes such as team rankings or player performance ratings. These models provide valuable insights into the interdependencies between different variables in the context of team dynamics and 
individual performances. However, sports data are often hierarchical and temporal, with interactions evolving over time. This introduces complexities that require adaptations of ordinal graphical models to account for these dependencies. For instance, \citet{nikolaou2024understanding} explore team collapse dynamics using probabilistic graphical models, while \citet{oh2015graphical} simulate basketball match outcomes with graphical models. Additionally, \citet{d2023bayesian} apply a Bayesian network approach to analyze basketball player performance, emphasizing the growing potential of graphical models in sports analytics. Future advancements in this area could benefit from integrating temporal and spatial effects, akin to the dynamic models used in epidemiology, to improve predictive accuracy and enhance decision-making in sports management and strategy.

The existing literature has thoroughly explored heterogeneity in Gaussian graphical models using both traditional non-penalized model-based clustering approaches \citep{scrucca2023mclust} and penalized likelihood methods \citep{ruan2011regularized, gao2016estimation, hao2018simultaneous}. However, \citet{haslbeck2023impact} investigated the application of Gaussian mixture models to ordinal rather than continuous data and found that, even with large sample sizes, the estimated component parameters exhibited persistent bias.
Beyond graphical models, clustering techniques tailored for ordinal data have also been developed. For instance, \citet{jacques2018model} proposed a parsimonious latent block model based on the Binary Ordinal Search distribution for co-clustering, while \citet{corneli2020co} introduced a generative model that accommodates sparse ordinal arrays by leveraging latent continuous variables. More recently, \citet{ernst2025ordinal} developed a model-based clustering approach in which the component distributions follow a beta–binomial specification.
Despite these contributions, few studies have addressed the simultaneous estimation of heterogeneous conditional relationships in ordinal data. 
A notable exception is \citet{lee2022estimating}, who introduced finite mixtures of ordinal graphical models to capture heterogeneous conditional dependence structures. This approach estimates both the mixing proportions and heterogeneous precision matrices via maximum penalized likelihood estimation. However, a critical limitation persists: the absence of a robust method for determining the optimal number of mixture components. This gap highlights the need for further methodological advancements to resolve model selection challenges in ordinal graphical models, ensuring both flexibility in capturing heterogeneity and parsimony in model complexity.

\subsection{Contributions}
We propose MFM-PGM, an innovative nonparametric Bayesian framework for modeling probit graphical models, leveraging the Mixture of Finite Mixtures (MFM) framework proposed by \citep{miller2018mixture}. This approach represents each cluster with a probit graphical model while allowing the number of clusters to vary flexibly. By integrating the random variation in the number of clusters with the model structure, we simultaneously estimate both the cluster configurations and the number of clusters, addressing key challenges in model selection for heterogeneous ordinal data. To implement this framework, we develop a computationally efficient Gibbs sampling algorithm tailored for full Bayesian inference. This algorithm estimates key parameters, including the number of clusters, mixture proportions, and precision matrices for the probit graphical models. Additionally, we incorporate advanced techniques to improve convergence and scalability, ensuring the model can handle large, complex datasets efficiently. The algorithm's design accommodates the unique challenges posed by ordinal data, such as the presence of latent variables and computational complexity, making it a robust solution for practical applications.

In the domain of sports analytics, this framework provides a powerful tool for uncovering hidden structures and interactions in ordinal datasets, which are common in sports research. Examples include analyzing within game player performance. By capturing heterogeneity within subgroups, such as players with distinct roles, our approach offers valuable insights into the underlying structure of sports data. These insights can inform performance evaluation, strategic decision-making, and personalized training programs.

Overall, our contributions include a novel nonparametric Bayesian framework for ordinal data modeling, an efficient computational algorithm for Bayesian inference, and impactful applications in sports analytics. This work addresses key methodological gaps, advancing both theory and practical applications, and lays the foundation for future exploration of ordinal graphical models in diverse domains.


\section{Methodology}
\subsection{Ordinal Probit Graphical Model}

We consider the probit graphical model used in \citet{guo2015graphical,lee2022estimating}. Suppose we have \( p \) ordinal random variables, \( X_1, \ldots, X_p \), where each \( X_j \) takes values in \( \{1, 2, \ldots, K_j\} \) for some integer \( K_j \), representing the number of ordinal levels in variable \( j \). We assume the existence of latent variables \( Z_1, \ldots, Z_p \) that follow a joint Gaussian distribution with mean vector \( \mathbf{\mu} \in \mathbb{R}^p \) and covariance matrix \( \mathbf\Sigma \). The observed variable \( X_j \) is derived by discretizing its latent counterpart \( Z_j \) based on a set of ordered thresholds, 
\(
-\infty = \theta_0^{(j)} < \theta_1^{(j)} < \cdots < \theta_{K_j-1}^{(j)} < \theta_{K_j}^{(j)} = +\infty.
\) 
The mapping follows, \(\mathrm{Pr}(X_{j}=k) = \mathrm{Pr}(\theta_{k-1}^{(j)} \leq Z_{j} < \theta_{k}^{(j)})\).

Define the precision matrix \( \mathbf{\Omega} = \mathbf{\Sigma}^{-1} \), the threshold set 
\(
\mathbf{\Theta} = \{ \theta_k^{(j)} : j = 1, \dots, p; k = 0, \dots, K_j \},
\)
and let \( \mathbf{X} = (X_1, \dots, X_p)' \) and \( \mathbf{Z} = (Z_1, \dots, Z_p)' \). The hypercube associated with \( \mathbf{X} \) and \( \mathbf{\Theta} \) is 
\(
C( \mathbf{X}, \mathbf{\Theta} ) = \prod_{j=1}^{p} [\theta_{X_j-1}^{(j)},\theta_{X_j}^{(j)}).
\)
The joint density function of \( (\mathbf{X}, \mathbf{Z}) \) is 
\begin{equation}\label{eqn-2} 
\begin{aligned}
  f_{\mathbf{X,Z}}(\mathbf{x},\mathbf{z}; \mathbf{\mu}, \mathbf{\Omega}, \mathbf{\Theta}) &= f(\mathbf{z}; \mathbf{\mu}, \mathbf{\Omega}) \prod_{j=1}^{p} I(z_j \in [\theta_{x_{j}-1}^{(j)},\theta_{x_{j}}^{(j)})) \\
  &= \frac{(\det\mathbf{\Omega})^{1/2}}{(2\pi)^{p/2}} \exp\left(-\frac{1}{2} (\mathbf{z}-\mathbf{\mu})' \mathbf{\Omega} (\mathbf{z}-\mathbf{\mu})\right)\times I(\mathbf{z} \in C(\mathbf{x}, \mathbf{\Theta})),
\end{aligned}
\end{equation}
where \( I(\cdot) \) is the indicator function.

Instead of sampling from the joint posterior distribution of only \( (\mathbf{\Theta}, \mathbf{\mu}, \mathbf{\Omega}) \), we focus on the joint posterior distribution of \( (\mathbf{\Theta}, \mathbf{\mu}, \mathbf{\Omega}, \mathbf{Z}) \), following the framework of \citet{albert1993bayesian}. This strategy is especially effective in Bayesian modeling, as it enables tractable posterior computation in models with intractable likelihoods by introducing latent variable augmentation. And evaluating the posterior of \( (\mathbf{\Theta}, \mathbf{\mu}, \mathbf{\Omega}) \) requires integrating over the truncated regions $C(\mathbf{x}, \mathbf{\Theta})$, which is analytically intractable and computationally intensive. The joint posterior distribution is given by:
\begin{equation}\label{eqn-4-4}
\begin{aligned}
\pi(\mathbf{\Theta}, \mathbf{\mu}, \mathbf{\Omega}, \mathbf{z}_1, \dots, \mathbf{z}_N \mid \mathbf{x}_1, \dots, \mathbf{x}_N) 
\propto \pi(\mathbf{\Theta}, \mathbf{\mu}, \mathbf{\Omega})\times\prod_{i=1}^{N} f_{\mathbf{X,Z}}(\mathbf{x}_i, \mathbf{z}_i; \mathbf{\mu}, \mathbf{\Omega}, \mathbf{\Theta}).
\end{aligned}
\end{equation}
We employ a sampling-based approach, leveraging Markov chain Monte Carlo (MCMC) methods \citep{chib1998analysis}, to explore and summarize the posterior distribution.

\subsection{Clustered Probit Graphical Model}

We extend the probit graphical model to a clustered framework. Suppose we have \( p \) ordinal random variables \( X_1, \dots, X_p \), where \( X_j \) takes values in \( \{1,2,\dots,K_j\} \). We assume the existence of latent variables \( Z_1,\ldots,Z_p \) following a Gaussian distribution with mean \( \mu_{c_i} \in \mathbb{R}^p \) and covariance \( \Sigma_{c_i} \), where \( c_i \in \{1,\dots,K\} \) represents cluster labels. Let \( \Omega_{c_i} = \Sigma_{c_i}^{-1} \) be the precision matrix. Let \( G = (V, E) \) be an undirected graph, where \( V \) contains \( p \) nodes and \( E \) describes conditional independence relationships. A Gaussian graphical model for \( \mathbf{Z} = (Z_1,\dots,Z_p)' \sim \mathcal{N}(\mathbf{\mu}_{c_i}, \mathbf{\Omega}_{c_i}^{-1}) \) is represented by \( G_{c_i} = (V, E_{c_i}) \), where variables \( Z_i \) and \( Z_j \) are independent given all others if and only if there is no edge \( (i,j) \) in \( E_{c_i} \). In this case, \( (\mathbf{\Omega}_{c_i})_{ij} = 0 \). Each observed variable \( X_j \) is discretized from its latent counterpart \( Z_j \) based on ordered thresholds. The joint density function of \( (\mathbf{X,Z}) \) follows:

\begin{equation}\label{eqn-2-2} 
\begin{aligned}
  & f_{\mathbf{X,Z}}(\mathbf{x}, \mathbf{z}; \mathbf{\mu}_{c_i}, \mathbf{\Omega}_{c_i}, G_{c_i}, \mathbf{\Theta}) \\
  &\quad = f(\mathbf{z}; \mathbf{\mu}_{c_i}, \mathbf{\Omega}_{c_i}, G_{c_i}) 
    \prod_{j=1}^{p} I\bigl(z_j \in [\theta_{x_{j}-1}^{(j)}, \theta_{x_{j}}^{(j)})\bigr) \\[6pt]
  &\quad = \frac{(\det \mathbf{\Omega}_{c_i})^{1/2}}{(2\pi)^{p/2}} 
    \exp\left(-\frac{1}{2} (\mathbf{z} - \mathbf{\mu}_{c_i})' \mathbf{\Omega}_{c_i} (\mathbf{z} - \mathbf{\mu}_{c_i})\right) 
    I\bigl(\mathbf{z} \in C(\mathbf{x}, \mathbf{\Theta})\bigr).
\end{aligned}
\end{equation}

Define \( \mathbf{\mu} = \{\mathbf{\mu}_{c_i}\}_{i=1}^n \), \( \mathbf{\Omega} = \{\mathbf{\Omega}_{c_i}\}_{i=1}^n \), and \( \mathbf{G} = \{G_{c_i}\}_{i=1}^n \). The density function of \( \mathbf{X} \) depends on \( \mathbf{\Theta}, \mathbf{\mu}_{c_i}, \mathbf{\Omega}_{c_i}, G_{c_i} \), and the latent variable \( \mathbf{Z} \). 
We focus on the joint posterior distribution:
\begin{equation}\label{eqn-4}
\begin{aligned}
\pi(\mathbf{\Theta},\mathbf{\mu},\mathbf{\Omega},\mathbf{G},\mathbf{z}_1,...,\mathbf{z}_N \mid \mathbf{x}_1,...,\mathbf{x}_N) 
\propto
\pi(\mathbf{\Theta},\mathbf{\mu},\mathbf{\Omega},\mathbf{G})\prod_{i=1}^N f_{\mathbf{X},\mathbf{Z}}(\mathbf{x}_i,\mathbf{z}_i;\mathbf{\mu}_{c_i},\mathbf{\Omega}_{c_i},G_{c_i},\mathbf{\Theta}).
\end{aligned}
\end{equation}
\subsection{Mixture of Finite Mixtures}
To flexibly model the cluster structure, we consider a nonparametric Bayesian approach. A popular prior for clustering is the **Chinese Restaurant Process (CRP)**, which defines the cluster assignment \( c_i, i = 2, \dots, N \) through the following conditional distribution \citep{ferguson1973bayesian}:

\begin{equation}
P(c_i = c \mid c_1, \dots, c_{i-1}) =
\begin{cases}
\frac{|c|}{i-1+\gamma}, & \text{if } c \text{ is an existing cluster,} \\
\frac{\gamma}{i-1+\gamma}, & \text{if } c \text{ is a new cluster,}
\end{cases}
\end{equation}
where \( |c| \) denotes the current size of cluster \( c \), and \( \gamma \) is a concentration parameter controlling the probability of creating new clusters. The CRP has the attractive property of simultaneously estimating the number of clusters and their configuration. However, recent studies \citep{miller2013simple} have shown that CRP often produces extraneous clusters in the posterior distribution, leading to inconsistent estimation of the true number of clusters even as the sample size increases to infinity. This issue arises because CRP favors small clusters due to its rich-get-richer property.
To address this inconsistency, the Mixture of Finite Mixtures (MFM) model has been proposed as a refinement of the CRP \citep{miller2018mixture}. Unlike CRP, MFM assumes that the number of clusters is a random variable \( K \) with a proper probability mass function \( p(\cdot) \), leading to a more stable estimation of the true number of clusters. The model is specified as:
\begin{equation}    
\begin{aligned}
K &\sim p(\cdot), \quad (\pi_{1},\ldots,\pi_{K}) \mid K \sim \mathrm{Dir}(\gamma,\ldots,\gamma), \\
c_{i} \mid K, \pi &\sim \sum_{h=1}^{K} \pi_{h} \delta_{h}, \quad i=1,\ldots,N,
\end{aligned}
\end{equation}
where \( \delta_h \) is a point-mass at \( h \), and \( \gamma \) is the concentration parameter of the Dirichlet Process. Compared to CRP, MFM slows down the creation of new clusters by a factor of \( V_N(t+1)/V_N(t) \), which automatically prunes tiny, extraneous clusters in a data-driven manner.
The coefficient \( V_N(t) \) is precomputed as:
\begin{equation}
V_N(t) = \sum_{k=1}^{+\infty} \frac{k_{(t)}}{(\gamma k)^{(N)}} p(k),
\end{equation}
where \( k_{(t)} = k(k-1) \dots (k-t+1) \) and \( (\gamma k)^{(N)} = \gamma k (\gamma k+1) \dots (\gamma k+N-1) \). The cluster assignment process in MFM can be expressed in a Pólya urn scheme, similar to CRP:
\begin{equation}
\begin{aligned}
P(c_{i} = c \mid c_{1}, \dots, c_{i-1}) 
\propto
\begin{cases}
|c| + \gamma, 
& \text{if } c \text{ is an existing cluster,} \\[8pt]
\gamma \dfrac{V_{N}(t+1)}{V_{N}(t)}, 
&  \text{if } c \text{ is a new cluster,}
\end{cases}
\end{aligned}
\end{equation}
where \( t \) denotes the current number of clusters. By incorporating \textcolor{purple}{} MFM model, we achieve a more robust clustering framework that mitigates the inconsistency issue in CRP while retaining its flexibility. This allows for data-adaptive estimation of the number of clusters, avoiding the overestimation problem inherent in the CRP.

\subsection{Prior Distribution for Parameters}

In this section, we introduce priors for \( (\pmb{{\Theta},\mu,\Omega,G}) \), structured as follows:
\begin{equation}
    \pi(\pmb{{\Theta},\mu,\Omega,G}) = \pi(\pmb{\Theta}) \pi(\pmb{\mu} \mid \pmb{\Omega}) \pi(\pmb{\Omega} \mid \pmb{G}) \pi(\pmb{G}),
\end{equation}
where each component of \( \pmb{\mu} \), \( \pmb{\Omega} \), and \( \pmb{G} \) is assumed to be independently and identically distributed. For clarity, we will denote \( \mu \in \pmb{\mu} \), \( \Omega \in \pmb{\Omega} \), and \( G \in \pmb{G} \) in subsequent sections.

We assign a diffuse prior for \( \pmb{\Theta} \), as in \citet{albert1993bayesian}, such that 
\(
p(\theta_k^{(j)}) \propto 1,  \forall k,j
\). For the mean \( \mu \), we assume a normal prior conditional on the precision matrix \( \Omega \), 
\(
\mu \mid \Omega \sim \mathcal{N}(\mu_0, (a\Omega)^{-1})
\). 
Following \citet{fraley2007bayesian}, we set \( \mu_0 = 0 \) and use \( a = 0.01 \) in simulations.

The G-Wishart distribution \citep{roverato2002hyper,letac2007wishart} serves as the standard conjugate prior for the precision matrix \( \Omega \) of a Gaussian graphical model. Given the graph structure \( G \), we assume a conjugate G-Wishart prior, 
\(
\Omega \mid G \sim W_G(b, D)
\), 
with density 
\[
p(\Omega \mid G) = I_G^{-1}(b, D) |\Omega|^{(b-2)/2} \exp\left(-\frac{1}{2} \operatorname{tr}(D \Omega)\right),
\]
where \( b > 2 \) is the degrees of freedom, \( D \) is a symmetric positive definite matrix, and \( I_G(b,D) \) is a normalizing constant. In our simulations, we set \( b = 3 \) and \( D = I \). The choices of $b$ and $D$ represent weakly informative priors, selected following the settings proposed by \citet{Mohammadi2015Bayesian}.
For the graph structure \( G \), we consider a Bernoulli prior on each edge inclusion indicator \citep{mohammadi2019bdgraph} 
\[
p(G) = q^{|E|} (1 - q)^{\frac{p(p-1)}{2} - |E|} \propto \left( \frac{q}{1 - q} \right)^{|E|},
\]
where \( |E| \) represents the number of edges in the graph \( G \), and \( q \in (0,1) \) is the prior probability of an edge existing. Setting \( q = 0.5 \) results in a uniform prior over the space of all graphs, serving as a non-informative prior. To encourage sparsity, we penalize additional edges by choosing a smaller \( q \). In our simulations, we set \( q = 0.2 \) as \citet{Mohammadi2015Bayesian}. These priors allow us to integrate structural information while maintaining flexibility in model estimation.

\subsection{Full Hierarchical Model}
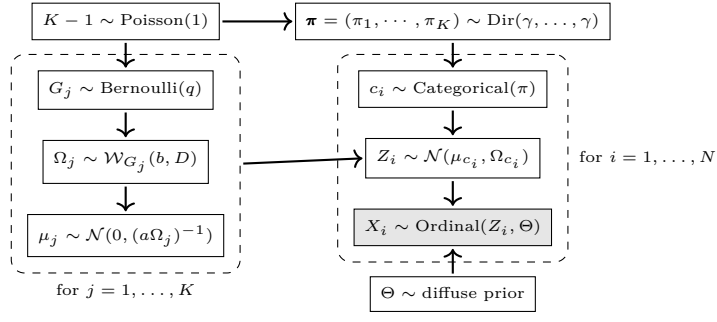
\begin{figure}[ht]
\centering
\begin{tikzpicture}[
  node distance=0.4cm and 0.4cm,
  font=\tiny,
  obs/.style={draw, rectangle, fill=gray!20, minimum width=2cm, align=center},
  latent/.style={draw, rectangle, minimum width=2cm, align=center},
  plate/.style={draw, dashed, inner sep=6pt, rectangle, rounded corners=5pt},
  arrow/.style={->, thick, shorten >=1pt, shorten <=1pt}
]

\node[latent] (Kprior) at (0,0) {$K-1 \sim \mathrm{Poisson}(1)$};
\node[latent, below=of Kprior] (G) {$G_j \sim \mathrm{Bernoulli}(q)$};
\node[latent, below=of G] (Omega) {$\Omega_j \sim \mathcal{W}_{G_j}(b, D)$};
\node[latent, below=of Omega] (mu) {$\mu_j \sim \mathcal{N}(0, (a\Omega_j)^{-1})$};

\node[plate, fit=(G)(Omega)(mu), label=below:{for $j = 1,\ldots,K$}] (plateK) {};

\node[latent, right=1cm of Kprior] (pi) {$\pmb\pi=(\pi_1,\cdots,\pi_K) \sim \mathrm{Dir}(\gamma,\ldots,\gamma)$};
\node[latent, below=of pi] (ci) {$c_i \sim \mathrm{Categorical}(\pi)$};
\node[latent, below=of ci] (Zi) {$Z_i \sim \mathcal{N}(\mu_{c_i}, \Omega_{c_i})$};
\node[obs, below=of Zi] (Xi) {$X_i \sim \text{Ordinal}(Z_i, \Theta)$};

\node[plate, fit=(ci)(Zi)(Xi), label=right:{for $i = 1,\ldots,N$}] (plateN) {};

\node[latent, below=of Xi] (Theta) {$\Theta \sim \text{diffuse prior}$};

\draw[arrow] (Kprior) -- (G);
\draw[arrow] (G) -- (Omega);
\draw[arrow] (Omega) -- (mu);
\draw[arrow] (Kprior.east) -- (pi.west);
\draw[arrow] (pi) -- (ci);
\draw[arrow] (ci) -- (Zi);
\draw[arrow] (plateK.east) -- (Zi.west);
\draw[arrow] (Zi) -- (Xi);
\draw[arrow] (Theta) -- (Xi);

\end{tikzpicture}
\caption{Hierarchical model diagram of MFM-PGM.}
    \label{fig:model_diagram}
\end{figure}

To account for the potential heterogeneity in the ordinal
data,  we propose a Bayesian mixture model where each mixture component is represented by a probit graphical model. More specifically, suppose that there are a total number of $K$
clusters, with weights $\pi_1,...,\pi_K$,  and each mixture follows a
different probit graphical model. Then we adopt the mixture of finite mixtures (MFM) framework \citep{miller2018mixture} by assigning prior distributions on those unknown model
parameters. Hence, the hierarchical model can be expressed
as follows.
\begin{equation}\label{eq:hierachical}
\begin{aligned}
&K\sim p_{K}, p_{K} \mathrm{~is~a~p.m.f~on~}\mathbb{N}^{+}=\{1,2,\ldots\},\\
&\pi=(\pi_{1},\ldots,\pi_{K})|K\sim\mathrm{Dir}(\gamma,\ldots,\gamma)\\
&p(c_{i}=j|\pi,K)=\pi_{j} \mathrm{~for~every~}  i=1,\ldots,N, \mathrm{and~}
j=1,\ldots,K,\\
&\mu_j{\sim} N(0,(a\Omega_j)^{-1})\mathrm{~for~every~}  j=1,\ldots,K,\\
&\Omega_1,...,\Omega_K\stackrel{\mathrm{iid}} {\sim}\mathrm{W}_G(b, D) , \\
&G_1,...,G_K \stackrel{\mathrm{iid}}{\sim}Bernoulli(q),\\
&Z_i\sim N(\mu_{c_{i}},\Omega_{c_{i}}),\\
&\Theta\sim \mathrm{a~ diffuse~ prior,}\\
&X_i\stackrel{}{\sim}\mathcal \mathrm{Probit~Graphical}(\mu_{c_{i}},\Omega_{c_{i}},G_{c_{i}},Z_i,\Theta),
\end{aligned}
\end{equation}
where $c_1,...,c_N$ are cluster membership indicators that take
values in $\{1, ... , K\}$ for each observation $X_i$.
 A default specification sets $p_K$ to be a $Poisson(1)$ distribution truncated to positive integers, and sets 
$\gamma=1$, following the guidance of \citep{miller2018mixture}.
 The diagram relationship is shown in Figure \ref{fig:model_diagram}.

\subsection{Theoretical Results}
In this section, we study the theoretical properties of the proposed model in \eqref{eq:hierachical}.
Given a fixed number of clusters, we assume the parameter space \( \Theta^* \) is compact for all model parameters, including mixture weights and mean and precision of the latent process. The mixing measure \( G^* \) is expressed as \( G^* = \sum_{h=1}^K \pi_h \delta_{\theta_h} \), where \( \delta \) represents the point mass measure, and \( \theta_h = \{\mu_h, \Omega_h, G_h\} \) includes the mean and precision of latent process in cluster \( h \). Let $K_0$, $G_0$, $P_0$ be the true number of clusters, the true mixing measure, and the corresponding probability measure, respectively. Then the following theorem establishes the posterior consistency and contraction rate for the cluster number $K$ and mixing measure $G^*$. It is based on the general results for Bayesian mixture models in \citet{guha2021posterior}.  
\begin{theorem}
\label{thm1}
Assume \( \Pi_n(\cdot | \mathbf{X}_1, \ldots, \mathbf{X}_N) \) denotes the posterior distribution from a random sample \( \mathbf{X}_1, \ldots, \mathbf{X}_N \) under a graph-based hierarchical model. If the parameters are confined to a compact space \( \Theta^* \), then:
\[
\Pi_n(K = K_0 | \mathbf{X}_1, \ldots, \mathbf{X}_n) \rightarrow 1,
\]
\[
\Pi_n(W(G^*, G_0) \preccurlyeq (\log N / N)^{{1/2}} | \mathbf{X}_1, \ldots, \mathbf{X}_N) \rightarrow 1,
\]
almost surely under \( P_0 \) as \( N \rightarrow \infty \), where \( W \) denotes the Wasserstein distance, reflecting the graph structure's influence on the mixing measure.
\end{theorem} 

To establish this theorem, we need to verify conditions (P.1)--(P.4) as detailed in \citet{guha2021posterior}:

\begin{enumerate}
    \item \textbf{Condition (P.1) - Compactness and Identifiability of the Model}: We restrict our parameters of interest to a compact space \( \Theta^* \). The multivariate normal distribution, graph Wishart distribution, and Bernoulli distribution are each first-order differentiable, ensuring the identifiability of the model.
    
    \item \textbf{Condition (P.2) - Nonzero Priors}: We assign a nonzero prior on all parameters, ensuring that each parameter contributes to the posterior distribution.
    
    \item \textbf{Condition (P.3) - Distributional Adequacy}: The multivariate normal, graph Wishart distribution, and Bernoulli distribution each satisfies this condition, providing the necessary theoretical support for our model's assumptions.
    
    \item \textbf{Condition (P.4) - Regularized Prior on $K$}: A tailored distribution, such as a truncated Poisson, is used to model the number of clusters.
\end{enumerate}

Theorem \ref{thm1} demonstrates that our proposed model is capable of correctly identifying the unknown number of clusters and the latent clustering structure, with the posterior probability approaching unity as the sample size increases. The requirement for a compact parameter space \( \bm{\Theta^*} \) is standard in Bayesian nonparametrics, as discussed in \citet {guha2021posterior}. This condition is practically relevant, as model parameters are expected to reside within a predefined range. For instance, it is reasonable to assume that mixture weights exceed a minimal threshold, such as 0.001\%, to yield meaningful clustering results.

\section{Bayesian Inference}
\subsection{ MCMC Algorithm }
We present a Gibbs sampler that enables efficient Bayesian inference for the proposed model. 
By exploiting the conditional conjugacy property in model specification , we derive a  Gibbs sampler algorithm
for efficient Bayesian inference. Detailed derivations of the full
conditionals are provided in  Section S1 of the supplementary materials.

 Based on the Algorithm 2 of \citet{neal2000markov}, we obtain the
following proposition that provides the full conditional distribution of $c_i,, i = 1,...,N$ while collapsing the number of clusters ${K}$.
\begin{proposition}
    The full conditional distributions $P(c_i|c_{-i},\cdot)$ is given by
$$P(c_i=k|c_{-i},\cdot)\propto\begin{cases}(N_{i,k}+\gamma)p(z_i|\mu_k,\Omega_k)&\text{at an existing cluster } k\\
\gamma \frac{V_N(t+1)}{V_N(t)} m(Z_i)&\text{if enter a new cluster }k\end{cases},$$
where $m(Z_i)={\int P(G)\frac{I_G(b+1,D+B-A)}{I_G(b,D)}dG} \frac{1}{(2\pi)^{p/2}} {\bigg(\frac{a}{a+1}\bigg)}^{p/2}$, $A=(a+1)^{-1}(z_i+a\mu_0)(z_i+a\mu_0)'$ and $B=z_iz_i'+ a\mu_0\mu_0'$, $N_{-i,k}$ { is the number of }$c_j${ for }$j\neq i${ that are equal to }$k$, $\gamma$ is a parameter of the Dirichlet distribution, $t$  refers to the cardinality of the existing clusters of the set
$\{1, 2, ... , N\} \backslash {i}$.

We summarize the  Gibbs sampling procedure for the  model in Algorithm \ref{a2}.
\begin{algorithm} 
	\caption{:  Gibbs sampling for MFM-PGM}     
	 \label{a2}       
	\begin{algorithmic}[1] 
	\State Initialize the number of groups $K$, the group assignment $\{c_i\}_{i=1}^N$
and the parameters $(\pmb{\Theta},\{\mu_{c_i},\Omega_{c_i},G_{c_i}\}_{i=1}^N)$.
\For { iter from 1 to M}
\For{ i from 1 to N}
\State Update $c_i$ conditional on $\pmb{\mu,\Omega,Z}$ for each observaion $i$ by the conditional distribution  in Proposition 3.1.
\State If updated $c_i$ belongs to existing groups, then $K$ does
not change; otherwise, if updated $c_i$ forms a new group, them
$K:=K+1.$
\EndFor
\For{$c$ from 1 to $K$}
\State Update $(\mu_c,\Omega_c,G_c)$ from the posterior distribution based on the  all the data points currently associated with cluster $c$.
\EndFor
\For{ i from 1 to N}
\State Update $z_i$ from $\mathcal{N}(\mu_{c_i},\Omega_{c_i})$, truncated to  $C(x_i,\pmb{\Theta})$.
\EndFor
\State Update $\{\theta_k^{(j)}\}_{k,j}$ from  a uniform distribution.
 \EndFor   
    
   \end{algorithmic} 
\end{algorithm}
    
\end{proposition}

\subsection{Post MCMC Inference}

To summarize the posterior samples that may have different
clustering structure (e.g., different number of clusters), we adopt Dahl’s method \citep{dahl2006model} to select
the best post burn-in iteration with the least squares criterion.
The estimate output by the best post burn-in iteration is then
taken as the final post MCMC estimator. 
Specifically, we take advantage of the co-membership matrix
to help us select the best post burn-in iteration. Define the comembership matrix as  ${\mathbf{B}}=(b_{ij})\in\mathbb{R}^{n\times n}$, where $b_{ij}=I(c_i=c_{j}).$ Therefore, the $(i,j)$th element of $\mathbf{B}$ denotes whether the $i$th observation is in the same cluster with the $j$th observation. We then choose the best post burn-in
iteration as $$m_{b}=\mathrm{argmin}_{1\leq m\leq M}\left\|\mathbf{B}^{(m)}-\overline{\mathbf{B}}\right\|_{F}^{2},$$
where $\mathbf{B}^{(m)}$ is the estimated co-membership matrix in the $m$th post burn-in iteration and $\overline{\mathbf{B}}=M^{-1}\sum_{m=1}^{M}\mathbf{B}^{(m)}.$ As a consequence, the best post burn-in iteration is selected by the closest $\mathbf{B}^{(m_b)}$ to the mean grouping co-membership matrix $\overline{\mathbf{B}}.$ The number of groups is then determined as $K^{(m_b)}.$ Accordingly, the post MCMC estimations of the parameters and memberships are given by $m_b$th iterative sampling.

\section{Simulation}
To demonstrate the finite sample performance of our proposed
method, we conduct a number of numerical studies in this
section. Our goal is to sample from the posterior distribution of the unknown
parameters $K$, $(c_1,...,c_n)\in\{1,\ldots K\}$, $\pmb{\Theta}$,  $\pmb{\mu}=(\mu_1,\ldots,\mu_K)$, $\pmb{\Omega}=(\Omega_1,\ldots,\Omega_K)$ and $\pmb{G}=(G_1,\ldots,G_K)$. Of these, we are most concerned about $K$, $(c_1,...,c_n)$ and $\pmb{G}=(G_1,\ldots,G_K)$. 

\subsection{Simulation Setting}
We consider 
different factors: 
the number of samples for each cluster $(n=N/K=100,200)$, and the number of ordinal variables  $(p=10,15)$,  in which the graph structures is Independent Graph/Neighbor Chain Graph/ Modified Neighbor Chain Graph.  The first  graph structure is
constructed with 1’s on the main diagonal,  and 0 at
all other entries. The first neighbor chain graph structure is constructed with 1’s on the main
diagonal, 0.5’s on the sub-diagonal and super-diagonal and 0 at
all other entries. The second neighbor chain graph structure is constructed with 1’s on the main
diagonal, 0.5’s on the sub-diagonal and super-diagonal  but only if the column index is even,   0.25’s on the diagonal that lies
directly below and to the left of the sub-diagonal and on the diagonal that lies directly above and
to the right of the super-diagonal,  again restricted to columns whose index is even.
After the  graph structures are obtained, we take the inverse
 and re-scale them to create the covariance matrices, where the diagonal entries are 1’s.

Regarding the data generation procedure, we first generate the latent continuous data in the
following way: each cluster is characterized by a unique mean vector $m_k$,  where each component is sampled independently from the set $\{0,-1,1\}$ without replacement. 
When the number of latent clusters $K=3$, we use the three types of graphs described above. Multivariate normal samples are generated using these parameters. The case of K=2 is presented  in Section S2.1 of the supplementary materials. We further evaluate the performance of our method for K=5, with details and results provided in Section S2.2 of the supplementary materials.

Next, we obtain observed ordinal data following a similar procedure in \citet{lee2022estimating}. We simulate $X=(X_1,\ldots,X_p)^{\prime}$,where $X_j=\sum_{l=1}^{K_j-1}$ $1(Z_j\geq\theta_l^{(j)})$ for $j=1,\ldots,p.$ Here,the sequence of thresholds $\theta_l^{^{(j)}}$ are drawn uniformly on $[\Phi^{-1}({(l-0.5)\frac{1}{K_j}}),\Phi^{-1}((l+0.5)\frac{1}{K_j})]$ for $l=1,\ldots,K_j-1.$ And here we let $K_j=3,\forall j$. This procedure guarantees the randomness of the simulated thresholds and ensures that the difference of the number of samples at each level is not too large. We generate a total of 100 datasets under these conditions and conduct 2000 MCMC iterations for each dataset, treating the first 1000 iterations as the burn-in phase.

To evaluate the effect of skewed cluster sizes, we conducted an additional simulation study using the same setting as the $K=3$ scenario, but with an imbalanced cluster size distribution: 50, 100, and 200 samples, respectively.

\subsection{Performance Measurements}
 We consider the following
measurements to evaluate the finite sample performance. First, we consider probability of choosing the correct number of clusters,  and we employ the root mean square error (RMSE) to evaluate the
estimation accuracy. The RMSE for adjacency graph $G$ is calculated as
$$\mathrm{RMSE}_{{G}}=\left\{(Rn)^{-1}\sum_{r=1}^{R}\sum_{i=1}^{n}\left\|\hat{G}^{(r)}_{\widehat{z}_{i}^{(r)}}-G_{z_{i}}\right\|_0^{2}\right\}^{1/2}.$$

 To measure the similarity between the estimated group memberships and the true group memberships, we use the Adjusted
Rand Index (ARI) (Rand 1971; Hubert and Arabie 1985). Define
$\{z_i : 1 \leq i \leq n\}$ as the set of the true group memberships. Then
ARI is defined as
$$\mathrm{ARI}^{(r)}=\frac{\mathrm{RI}^{(r)}-E\Big(\mathrm{RI}^{(r)}\Big)}{\max\left(\mathrm{RI}^{(r)}\right)-E\Big(\mathrm{RI}^{(r)}\Big)},\quad\mathrm{RI}^{(r)}=\frac{a^{(r)}+b^{(r)}}{C_{N}^{2}},$$
where $a^{(r)} = \sum_{i,j}I(c_{i} = c_{j},\widehat{c}_{i}^{(r)} = \widehat{c}_{j}^{(r)})$,  and $b^{(r)} = \sum_{i,j}I(c_{i} \neq 
c_{j},\widehat{c}_{i}^{(r)} \neq \widehat{c}_{i}^{(r)})$.  Hence, ARI
measures the alignment level of two grouping results. A higher
ARI value implies the estimated group memberships are more
consistent with the true memberships. 

\subsection{Simulation Results}

{We compare our proposed method with 
PLE \citep{ruan2011regularized}, BOSClust \citep{jacques2018model}, OLBM \citep{corneli2020co} and Beta–binomial \citep{ernst2025ordinal}. To strengthen our comparison, we include an ablation study replacing the MFM prior with CRP prior. All methods are applied to ordinal data, and we additionally apply mclust \citep{scrucca2023mclust} to both the ordinal observations and their latent continuous data. }
Results for latent clusters $K=2$, $K=3$, $K=5$ and unbalanced data are detailed in Table S1 (supplementary materials), Table \ref{ta2}, Table S2 (supplementary materials), and Table \ref{ta-unba}, respectively. And BOSClust, OLBM and Beta–binomial's RMSEs are missing due to their lack of graphical model learning. We assess performance based on the accuracy of cluster number estimation, reporting the proportion of instances in which the true cluster number was correctly identified across 100 replicates. For $K=2$, $K=3$ and unbalanced data, our method successfully recovered the true number of clusters in 100\% of the cases, and for $K=5$, it achieved successful recovery in over 85\% of the replicates. These results demonstrate that our method achieves performance comparable to mclust applied to the underlying continuous data, while substantially outperforming all other competing methods.

Additionally, our approach consistently yields high Adjusted Rand Index (ARI) values, indicating strong agreement between the estimated and true group memberships. In contrast, PLE attains lower RMSE in the sparse setting by producing overly sparse graphs that overlook important variable dependencies. Refer to Section S2.1 of the supplementary materials for further details. 

\begin{longtable}{lllll}
    \caption{Simulation results for $K=3$.}\label{ta2} \\
    \hline
    ~ & ~ & Prob &  ARI  & RMSE  \\ \hline
    \endfirsthead
    
    \hline
    ~ & ~ & Prob &  ARI  & RMSE  \\ \hline
    \endhead
    
    \hline
    \endfoot
        n=100,p=10 & MFM-PGM & 1.00 (0.00) & 0.7383 (0.2127)  & 4.2462 \\ 
        ~ & PLE &0.00 (0.00) & 0.0779 (0.1706)&   3.4645\\
         ~&CRP& 0.32 (0.47)&0.6830 (0.1860) &4.4514 \\
         ~&BOSClust&0.38 (0.49) &0.4002 (0.2484) &--- \\
         ~&OLBM& 0.06 (0.24)&0.0146 (0.0713) &--- \\
         ~&mclust&0.00 (0.00)&0.4217 (0.0948)& 5.7020\\
         ~&Beta–binomial&0.49 (0.50)&0.7126 (0.0737)&---\\
~&mclust (con)& 0.97 (0.17)&0.9593 (0.0529)&8.8086\\
        \hline

        n=100,p=15  & MFM-PGM & 1.00 (0.00)&  0.7556 (0.2490)  & 6.2888 \\ 
        ~ & PLE &0.01 (0.10)& 0.3015 (0.2433)  &4.2434 \\
         ~&CRP&0.26 (0.44) &0.6990 (0.1982) &6.4928 \\
         ~&BOSClust&0.26 (0.44) &0.6256 (0.2237) &--- \\
         ~&OLBM&0.06 (0.24) &0.0096 (0.0451) &--- \\
         ~&mclust&0.01 (0.10)&0.5761 (0.1197)&4.6083\\
         ~&Beta–binomial&0.18 (0.39)&0.8005 (0.0504) &---\\
~&mclust (con)&0.97 (0.17)&0.9884 (0.0261)&13.8076\\
        \hline
        
        n=200,p=10 & MFM-PGM & 1.00 (0.00)& 0.7744 (0.2035)  &3.8717\\ 
        ~ & PLE & 0.06 (0.24)&  0.4108 (0.1704)  &3.4639  \\
         ~&CRP&0.44 (0.50) &0.7411 (0.1877) & 4.0170\\
         ~&BOSClust&0.32 (0.47) & 0.3926 (0.2487)&--- \\
         ~&OLBM& 0.08 (0.27)&0.0101 (0.0611) & ---\\
         ~&mclust&0.00 (0.00)&0.3932 (0.0814)&7.9911\\
         ~&Beta–binomial& 0.11 (0.31)&0.6977 (0.0616)&---\\
~&mclust (con)&0.96 (0.20)&0.9612 (0.0407)&8.7805\\
        \hline
        
        n=200,p=15 & MFM-PGM & 1.00 (0.00) & 0.8079 (0.2318) & 5.7164\\ 
        ~ & PLE & 0.86 (0.35)&0.8091 (0.1286)  & 4.2329\\
         ~&CRP&0.30 (0.46) & 0.7619 (0.1958)&5.9343 \\
         ~&BOSClust&0.30 (0.46) & 0.6105 (0.2430)&--- \\
         ~&OLBM& 0.04 (0.20)&0.0037 (0.0155) &--- \\
         ~&mclust&0.13 (0.34)&0.5759 (0.1682)& 8.4349\\
         ~&Beta–binomial&0.03 (0.17)&0.8027 (0.0468) &---\\
~&mclust (con)&1.00 (0.00)&0.9947 (0.0063)&13.7773\\
        \hline
\end{longtable}

\begin{longtable}{lllll}
    \caption{Simulation results for unbalanced data.}\label{ta-unba} \\
    \hline
    ~ & ~ & Prob &  ARI  & RMSE  \\ \hline
    \endfirsthead

    \hline
    ~ & ~ & Prob &  ARI  & RMSE  \\ \hline
    \endhead

    \hline
    \endfoot 
        p=10 & MFM-PGM & 1.00 (0.00) & 0.9131 (0.0473)  & 3.6463 \\ 
        ~ & PLE &0.00 (0.00) & 0.1807 (0.2838)  & 3.9254\\
~ & CRP &0.22 (0.42)  & 0.7066 (0.2267) & 4.2367\\
~&BOSClust&0.38 (0.49) &0.4317 (0.2389) &---\\
~&OLBM&0.04 (0.20) &0.0222 (0.0935) &---\\
~&mclust&0.02 (0.14)&0.3095 (0.0925)&7.0654\\
~&Beta–binomial&0.57 (0.50)&0.6738 (0.1241) &---\\
~&mclust (con)&0.90 (0.30)&0.9617 (0.0546)&8.6089\\
\hline

        p=15 & MFM-PGM & 1.00 (0.00)& 0.9790 (0.0170)  & 5.4312 \\ 
        ~ & PLE & 0.00 (0.00)& 0.4443 (0.3309)   & 4.7784\\ 
       ~ & CRP &0.28 (0.45) & 0.7930 (0.1671) &6.0642 \\ 
       ~&BOSClust&0.22 (0.42) &0.5703 (0.2501) &---\\
    ~&OLBM&0.00 (0.00)&0.0319 (0.1174) &---\\
    ~&mclust&0.01 (0.10)&0.3738 (0.0897)&5.1116\\
    ~&Beta–binomial&0.29 (0.46)&0.7834 (0.1266) &---\\
~&mclust (con)&0.96 (0.20)&0.9920 (0.0165)&13.6326\\
       \hline
    
\end{longtable}

\section{Empirical Case Study}

\subsection{Data Description}
In this section, we apply the proposed model to analyze the latent heterogeneity in real-world sports data from professional basketball players in the National Basketball Association (NBA). We focus on the 2017-2018 NBA season data from \url{NBAsavant.com}. The dataset consists of performance metrics and advanced basketball statistics for 536 NBA players, with 20 covariates capturing various aspects of their gameplay, including efficiency, shooting, rebounding, playmaking, defensive contributions, and overall impact. Details of these covariates are provided  in Section S3.1 of the supplementary materials. To facilitate analysis, each covariate is discretized into tertiles based on its quantiles and converted into an ordinal variable. Understanding the heterogeneity in player performance metrics is crucial for optimizing team-building strategies, refining game tactics, and enhancing overall franchise success in the NBA. Our analysis aims to uncover heterogeneous associations among these performance metrics, providing insights into the underlying structure of player capabilities and playing styles.

\subsection{Data Analysis}
To implement our method, we run an MCMC chain with a total of 6,000 iterations, discarding the first 3,000 draws as burn-in to ensure convergence. After applying Dahl’s method, we identify three distinct groups of players, consisting of 411, 112, and 13 individuals, respectively. These groups are referred to as Group 1, Group 2, and Group 3.

\begin{figure}

	\centering  
	\subfigbottomskip=2pt 
	\subfigcapskip=-5pt 
	\subfigure[The estimated  graphical model for Group 1]{
		\includegraphics[width=0.48\linewidth]{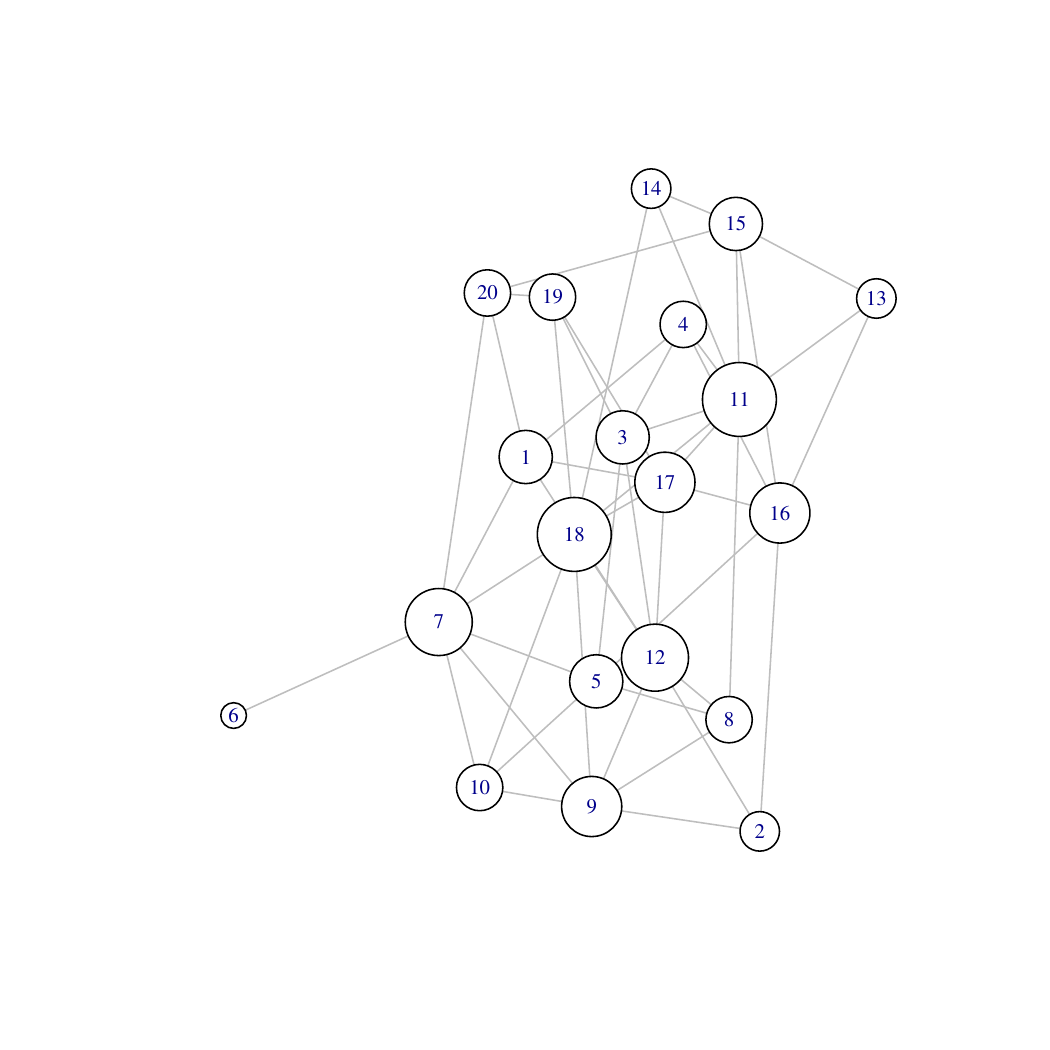}}
	\subfigure[The estimated  graphical model for Group 2]{
    \includegraphics[width=0.48\linewidth]{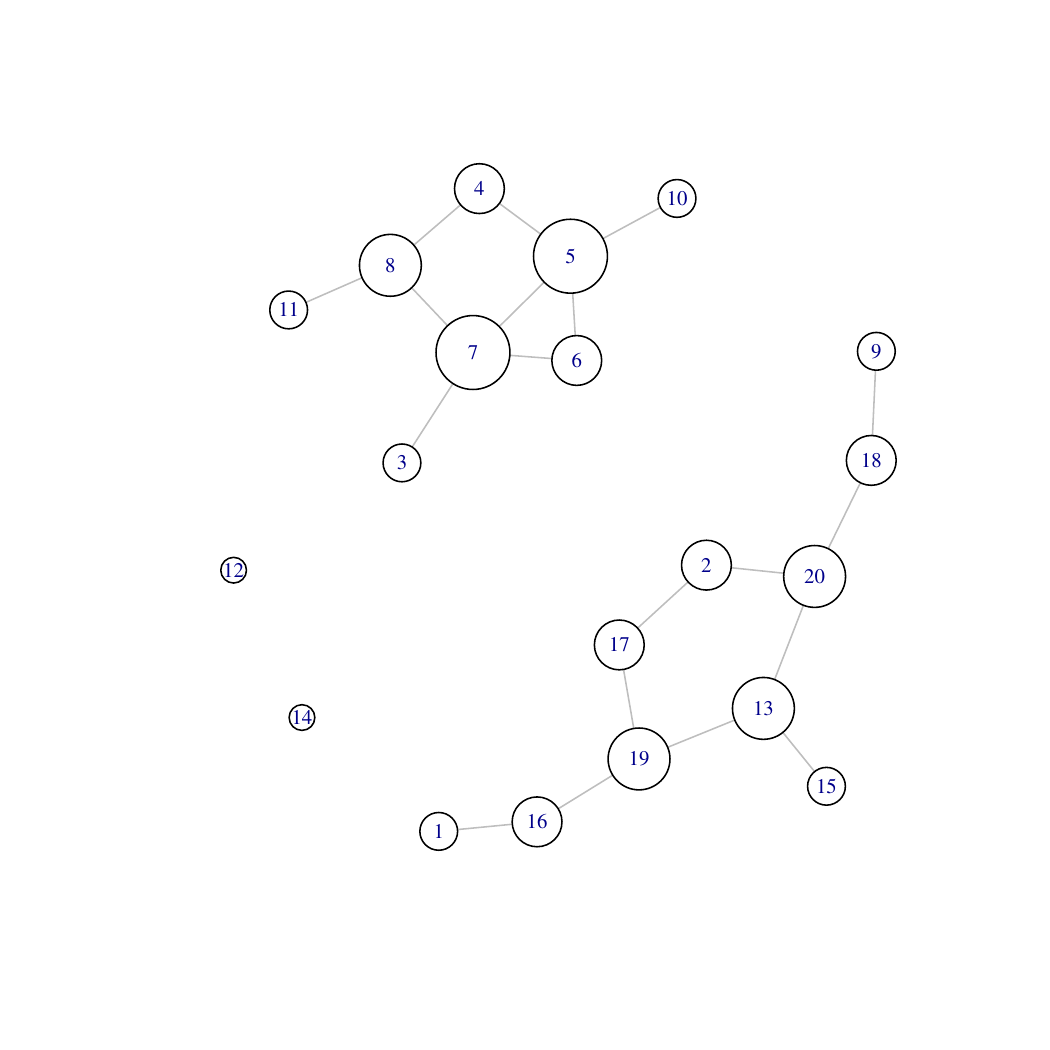}}
	  \\
	\subfigure[The estimated  graphical model for Group 3]{
		\includegraphics[width=0.48\linewidth]{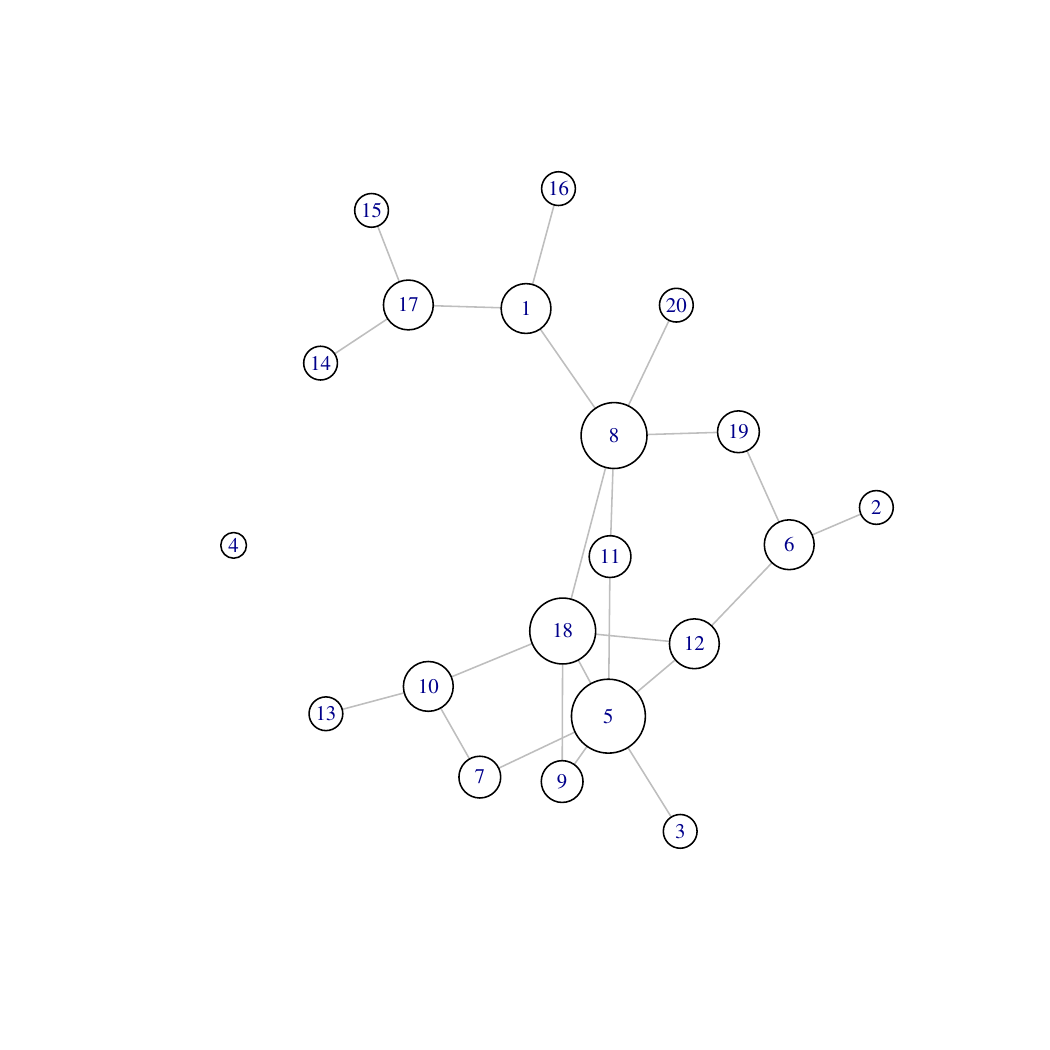}}
		\caption{The estimated ordinal graphical models. Nodes are scaled to degree.}
    \label{fig1}
\end{figure}

Table \ref{ta4} summarizes the descriptive network statistics for the three estimated groups. Our proposed method constructs a graphical model for each group, capturing the estimated group-level conditional dependencies among the 20 ordinal variables. The estimated edges of these graphical models are visualized in Figure \ref{fig1}, where node size is proportional to its degree (i.e., the number of edges connected to a node).
\begin{longtable}{llll}
    \caption{Summary of network statistics for three estimated groups.}\label{ta4} \\
    \hline
    Group ID & 1 & 2 & 3 \\ \hline
    \endfirsthead
    
    \hline
    Group ID & 1 & 2 & 3 \\ \hline
    \endhead
    
    \hline
    \endfoot
    
    N& 411 & 112 & 13\\
Maximum degree&8&4&6\\
Total degree&98&38&46\\
Average degree centrality&0.26&0.10&0.12\\
Average  betweenness centrality &8.65&4.90&17.60\\ \hline
\end{longtable}

Next, we compare the structural characteristics of Group 1, Group 2, and Group 3 based on their estimated graphical models. Our analysis reveals clear differences in the conditional dependence structures across the three groups, highlighting the heterogeneous relationships among the 20 performance metrics.
  
As shown in Figure \ref{fig1} and Table \ref{ta4}, the three estimated graphical models exhibit distinct network structures. 
Beyond overall network connectivity, the estimated graphs also differ in their hub nodes—variables with significantly more connections than average.
\begin{itemize}
    \item Group 1: The primary hub nodes include Turnover Percentage (Node 11), Defensive Box Plus/Minus (Node 18), Total Rebound Percentage (Node 7), Usage Percentage (Node 12), Steal Percentage (Node 9), Win Shares Per 48 Minutes (Node 16), and Offensive Box Plus/Minus (Node 17).
    \item Group 2: While Total Rebound Percentage (Node 7) remains a key hub, other prominent nodes are Offensive Rebound Percentage (Node 5), Assist Percentage (Node 8), Offensive Win Shares (Node 13), Box Plus/Minus (Node 19), and Value over Replacement Player (Node 20).
    \item Group 3: Hub nodes include those shared with Group 1 (Defensive Box Plus/Minus (Node 18), Usage Percentage (Node 12), Offensive Box Plus/Minus (Node 17)) and Group 2 (Offensive Rebound Percentage (Node 5), Assist Percentage (Node 8)), along with additional key metrics such as Player Efficiency Rating (Node 1), Defensive Rebound Percentage (Node 6), and Block Percentage (Node 10).
\end{itemize}
These differences in network structure and hub nodes underscore the heterogeneous dependencies among performance metrics across groups, revealing distinct patterns in player characteristics and playing styles.

\begin{figure}
    \centering
    \includegraphics[width=1\linewidth]{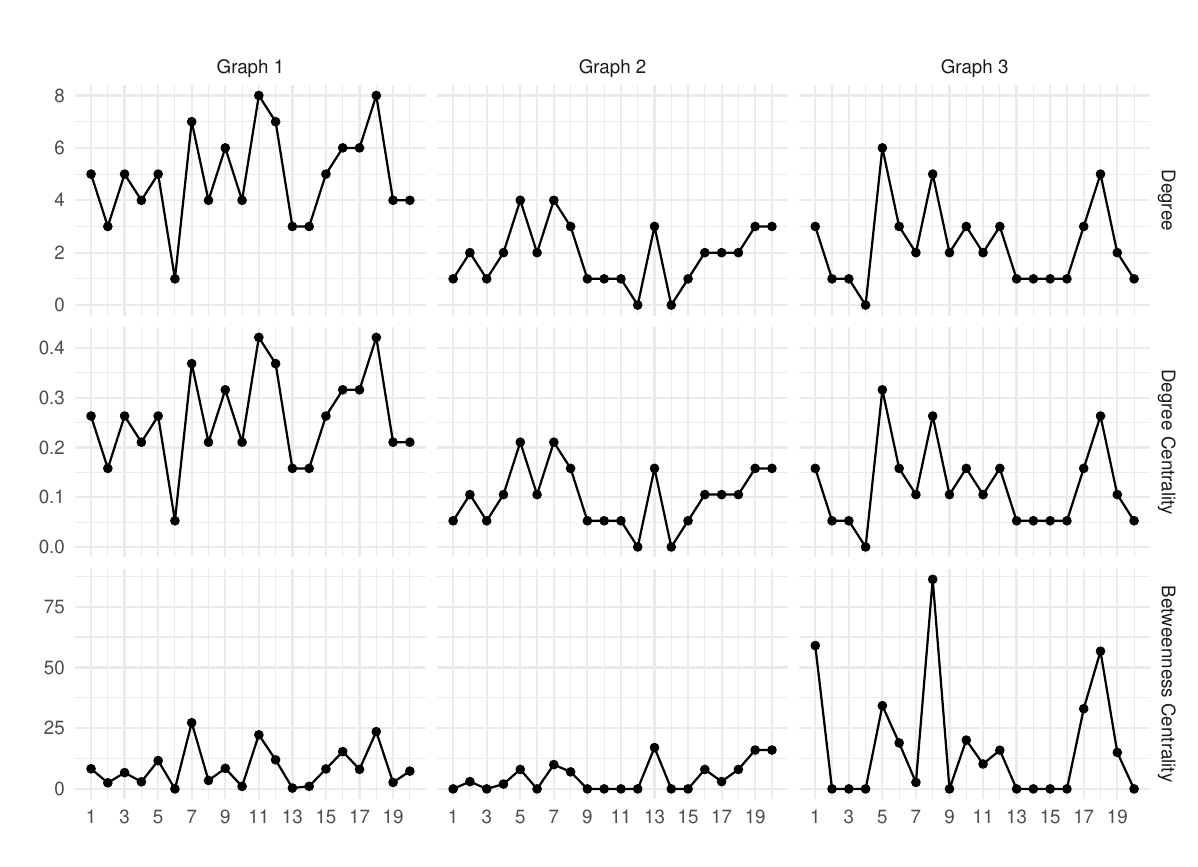}
        \caption{The degree and centrality plots of the three groups identified by MFM-PGM.}
    \label{fig:3}
\end{figure}

Figure \ref{fig:3} further highlights the differences in network structures among Groups 1, 2, and 3. The degree and centrality plots in Figure \ref{fig:3} illustrate these structural variations, where the x-axis represents node indices, and the y-axis denotes the corresponding network measures (i.e., degree, degree centrality, and betweenness centrality). The distinct patterns in these plots confirm the heterogeneity across the three groups. A full list of player names for each group is provided in Section S3.2 of the supplementary materials. The composition of these groups suggests meaningful distinctions in player roles and career trajectories:
\begin{itemize}
    \item Group 1: Established Stars and High-Impact Players. 
This group likely includes elite NBA players with consistent success, such as starters or key contributors to championship-contending teams. Notable names include 
Kevin Durant, Stephen Curry, and Kawhi Leonard, all of whom have multiple All-Star appearances, MVPs, and championship credentials. These players represent the top tier of talent, excelling in scoring, playmaking, and leadership.
\item Group 2: Role Players and Experienced Contributors. 
 This group comprises valuable rotation players, specialists, and veteran veterans who excel in supporting roles. Such as Patrick Beverley with his elite perimeter defense, Dorian Finney-Smith as a reliable stretch forward, and Udonis Haslem as a veteran locker room leader. While not primary stars, they make significant contributions through niche skills or invaluable experience. Their consistent impact within assigned roles complements core players and strengthens overall team depth.

\item Group 3: Adaptive Tactical Hubs. Defined by exceptional role switchability, players in this cluster dynamically adjust their on-court functions to address tactical demands. Their value stems from versatile gap-filling that optimizes system efficiency. Among stars, LeBron James  transformed into a de facto point center after Kyrie Irving’s departure, serving as both primary scorer and offensive initiator. Similarly, Giannis Antetokounmpo  leveraged unique physicality to fluidly toggle between rim-roller and switch defender. For role players,  Kyle Anderson operated as a 1-4 positional chameleon, excelling at both multi-position defense and secondary playmaking.
The group thrives on contextual elasticity with adaptation constituting their unifying trait.
\end{itemize}
The observed differences in network structures and player classifications provide valuable insights into how performance metrics correlate with different stages of an NBA player’s career.

In addition to the NBA data analysis, we analyze three real-world datasets to further demonstrate the practical utility of the proposed model in Section S4 of the supplementary materials. 
These include a dataset from the National Football League focusing on wide receivers across multiple seasons, a dataset from the China Health and Retirement Longitudinal Study capturing ordinal measures related to health and behavior among older adults, and a single-cell flow cytometry dataset examining signaling pathways under controlled stimulation.

\section{Discussion}
In this study, we develop a Bayesian heterogeneous probit graphical model to analyze ordinal data with complex dependencies, particularly in the context of sports analytics. The model incorporates both individual-level performance metrics, allowing for heterogeneous relationships across different player groups. By leveraging latent variable structures and clustering, our approach captures variations in player interactions and performance outcomes, enabling the model to account for different playing styles. Additionally, we explore how specific player attributes, such as scoring efficiency or defensive contributions, influence the overall team performance. Through comparisons with traditional methods and other competing models, we demonstrate the superiority of our approach in capturing these intricate relationships and improving the accuracy of player and team performance predictions.

The Bayesian heterogeneous probit graphical model overcomes key limitations of traditional sports analytics models, such as those relying on aggregated statistics. Unlike models that treat all players as homogeneous, our method allows for the estimation of individual player effects while accounting for interactions within teams and across games. This provides more granular insights into how different players contribute to team success, which is valuable for strategic decision-making in areas like player recruitment or in-game tactics. While detailed player-level data may require approval or be restricted, our model’s ability to integrate available demographic and performance data ensures it remains a powerful tool for both sports analytics and decision-making in competitive settings.

Future advancements in heterogeneous ordinal graphical models for sports could focus on several key areas. One promising direction is the application of variational inference techniques to approximate the posterior distributions of model parameters. This approach could provide a more scalable and efficient alternative to traditional inference methods, especially when dealing with large and complex datasets in sports analytics. Additionally, there is potential for incorporating temporal dependencies, which would enable the models to better reflect the dynamic dependent nature of team performance over the course of a game or season. Another avenue for improvement is extending these models to handle high-dimensional data, such as those involving numerous players and teams across multiple seasons. Integrating these models with other machine learning techniques, such as deep learning or reinforcement learning, could also enhance predictive accuracy and provide more actionable insights. Ultimately, these advancements will help uncover deeper relationships in sports performance data, improving decision-making in areas such as team selection, game strategy, and player development.

\phantomsection\label{disclosure-statement}
\bigskip

\begin{center}

{\large\bf Disclosure Statement}

\end{center}
The authors report there are no competing interests to declare.

\phantomsection
\bigskip

\begin{center}

{\large\bf Supplementary Materials}

\end{center}
Technical details of the posterior derivation, additional numerical results, and analyses of additional real datasets are provided in the Supplementary Materials.

  \bibliography{bibliography.bib}

\end{document}